\documentclass[sigconf]{acmart}
\AtBeginDocument{%
  \providecommand\BibTeX{{%
    Bib\TeX}}}

\copyrightyear{2025}
\acmYear{2025}
\setcopyright{cc}
\setcctype{by-sa}
\acmConference[MM '25]{Proceedings of the 33rd ACM International Conference on Multimedia}{October 27--31, 2025}{Dublin, Ireland}
\acmBooktitle{Proceedings of the 33rd ACM International Conference on Multimedia (MM '25), October 27--31, 2025, Dublin, Ireland}\acmDOI{10.1145/3746027.3755766}
\acmISBN{979-8-4007-2035-2/2025/10}

\title{Exploring Adapter Design Tradeoffs for Low Resource Music Generation}

\author{Atharva Mehta}
\authornote{Both authors contributed equally to this research.}
\email{atharva.mehta@mbzuai.ac.ae}
\affiliation{%
  \institution{Mohamed bin Zayed University of AI}
  \city{Abu Dhabi}
  \country{UAE}
}

\author{Shivam Chauhan}
\authornotemark[1]
\email{shivam.chauhan@presight.ai}
\affiliation{%
  \institution{Presight, G42 Company}
  \city{Abu Dhabi}
  \country{UAE}
}

\author{Monojit Choudhury}
\email{monojit.choudhury@mbzuai.ac.ae}
\affiliation{%
  \institution{Mohamed bin Zayed University of AI}
  \city{Abu Dhabi}
  \country{UAE}
}

\usepackage{multirow}
\usepackage{amsmath}
\usepackage{makecell} % Add this to your preamble
\usepackage{hhline} % Add this to your preamble
\usepackage{tabularx} % Add this line in the preamble
\usepackage{hyperref}
\usepackage{subcaption}  % Add this in your preamble
\usepackage{balance}

\begin{document}

\begin{abstract}
Fine-tuning large-scale music audio generation models, such as \textit{MusicGen} and \textit{Mustango}, is a computationally expensive process, often requiring updates to billions of parameters and, therefore, significant hardware resources. Parameter-Efficient Fine-Tuning (PEFT) techniques, particularly adapter-based methods, have emerged as a promising alternative, enabling adaptation with minimal trainable parameters while preserving model performance. However, the design choices for adapters, including their architecture, placement, and size, are numerous, and it is unclear which of these combinations would produce optimal adapters and why, for a given case of low-resource music genre. In this paper, we attempt to answer this question by studying various adapter configurations for two AI music models, \textit{MusicGen} and \textit{Mustango}, on two genres: Hindustani Classical and Turkish Makam music. 

Our findings reveal distinct trade-offs: convolution-based adapters excel in capturing fine-grained local musical details such as ornamentations and short melodic phrases, while transformer-based adapters better preserve long-range dependencies crucial for structured improvisation. Additionally, we analyze computational resource requirements across different adapter scales, demonstrating how mid-sized adapters (40M parameters) achieve an optimal balance between expressivity and quality. 
Furthermore, we find that \textit{Mustango}, a diffusion-based model, generates more diverse outputs with better adherence to the description in the input prompt while lacking in providing stability in notes, rhythm alignment, and aesthetics. Also, it is computationally intensive and requires significantly more time to train. In contrast, autoregressive models like \textit{MusicGen} offer faster training and are more efficient, and can produce better quality output in comparison, but have slightly higher redundancy in their generations. We release our datasets, models and training code in the following github repository: \href{https://github.com/atharva20038/ACMMM_Adapters/tree/main}{Github}.
\end{abstract}




\maketitle

\section{Introduction}\label{sec:introduction}
Research in music generation has advanced significantly with the emergence of large-scale generative models that can create high-quality compositions in a wide range of musical styles. \cite{c:24,c:23,c:22,agostinelli2023musiclm,melechovsky-etal-2024-mustango}. However, since they are trained on and optimized for data that mostly comes from the Western musical traditions, rich and diverse musical cultures such as Hindustani Classical \cite{jairazbhoy1971rāgs} and Turkish Makam music \cite{signell2008makam}, remain severely underrepresented\cite{mehta2024missingmelodiesaimusic}. Besides ethical concern around exclusion or disparate treatment of musical traditions, this imbalance limits the ability of music generation models to capture the full range of global musical expression \cite{10.1093/pnasnexus/pgae346,c:23}.

Furthermore, fine-tuning (i.e., updating all the parameters of) large music generation models such as \textit{MusicGen} \cite{c:23} and \textit{Mustango} \cite{melechovsky-etal-2024-mustango} demand large-scale computational infrastructure, making it difficult to scale, especially for low-resource genres \cite{aucouturier2003representing} with limited training data. To address this, researchers have turned to Parameter-Efficient Fine-Tuning (PEFT) \cite{houlsby2019parameter} techniques, which have gained popularity in NLP for enabling large pre-trained models to be adapted using lightweight modules such as adapters \cite{pfeiffer2020adapterhub,pfeiffer2020adapterfusion} or prompt tokens \cite{houlsby2019parameter,liu-etal-2024-alora}. These methods dramatically reduce computational costs by freezing the original model weights and only training a small number of additional parameters, making them ideal for domain adaptation under resource constraints.

Recent work by \citet{mehta2025music} demonstrates how Parameter-Efficient Fine-Tuning (PEFT) techniques can help adapt large music generation models to under-represented genres. However, their experiments reveal mixed results for the two models studied. Moreover, the authors do not explore different styles, positions and size of adapter configurations. In this work, we take a step toward answering these questions by systematically exploring the application of PEFT techniques (specifically, adapter-based) to two state-of-the-art music generation models \textit{MusicGen} and \textit{Mustango}, focusing on their adaptation to underrepresented genres. We analyze how the adapter architecture and placement affect musical quality, efficiency, and genre-specific expressiveness, providing detailed insights into the practical use of PEFT for culturally inclusive music generation, so that bringing low-resources genres to the manifold of AI music generation is practical and easily replicable.

\begin{figure*}[!t] \centering \includegraphics[width=1.01\linewidth,height=20cm, keepaspectratio]{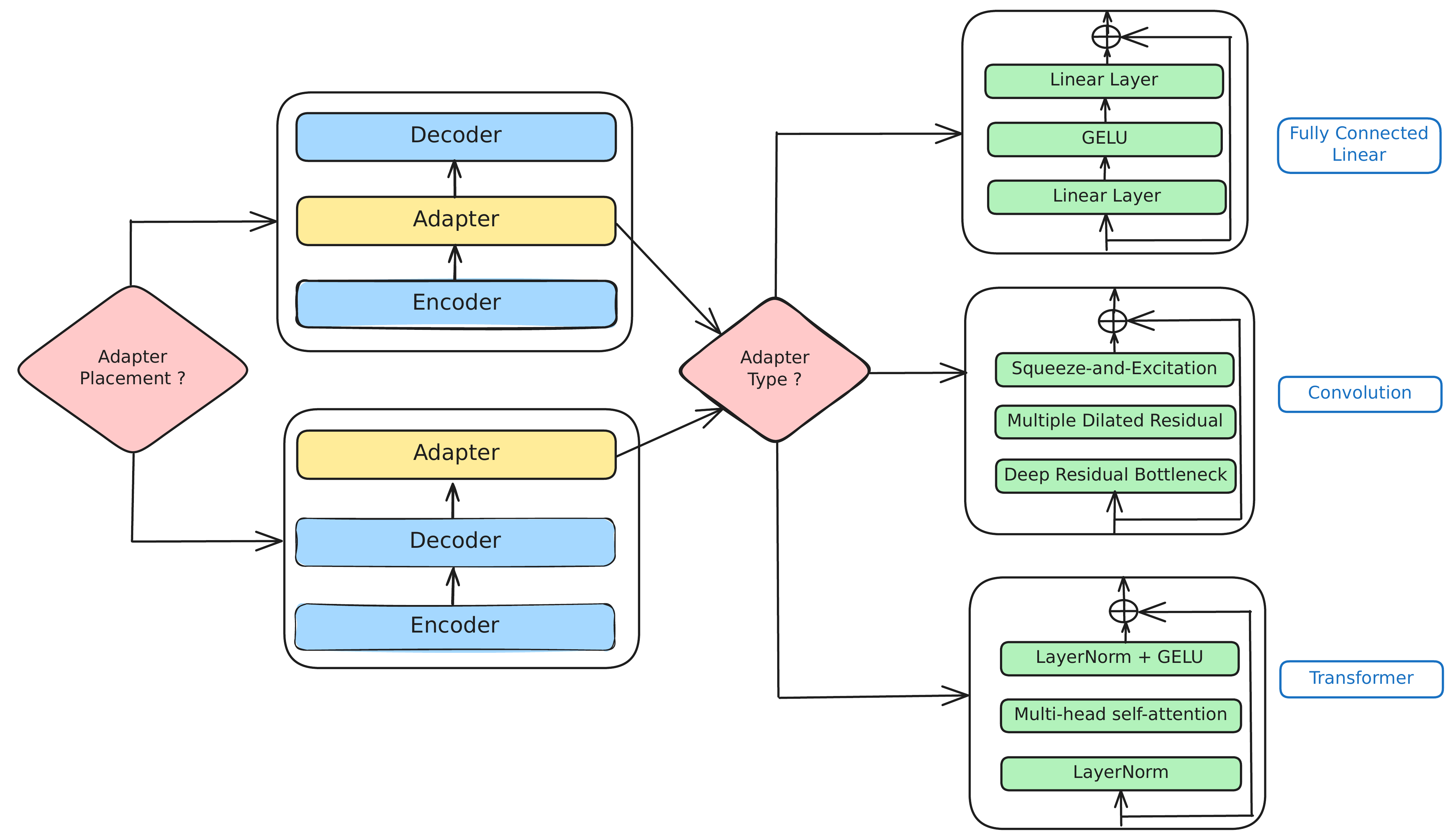} \caption{Adapter-based fine-tuning: Exploring different placements and architecture types (fully connected layers, Convolution, Transformer) in an encoder-decoder architecture.} \label{fig:architecture} \end{figure*}

 The contributions of our paper are threefold: 
\begin{enumerate}
\item We study the impact of adapter \textit{size}, \textit{placement} and \textit{architecture} in fine-tuning large generative music models for low-resource music genre:
\begin{itemize}
    \item \textit{Placement}: We show that placing adapters in the late layers of \textit{MusicGen} and \textit{Mustango} enhances generation quality while minimizing interference with core musical representations.
    \item \textit{Architecture}: We systematically compare fully connected layer-based, Convolution-based, and Transformer-based adapters, analyzing their impact on generation quality and computational efficiency.
    \item \textit{Size}: We show that there is an optimal adapter size for each model, given the size of the datasets and the base model, and larger-sized adapters hurt the performance.
\end{itemize}

    \item We conduct a large-scale evaluation of computational cost versus musical quality using objective metrics (Fréchet Audio Distance (FAD) \cite{Kilgour2019FrchetAD}, Fréchet Distance (FD) \cite{arabzadeh-clarke-2024-frechet}). Our results highlight optimal adapter sizes and architectures that achieve high-quality music generation while minimizing resource demands.
    \item We also extend adapter-based fine-tuning to Hindustani Classical and Turkish Makam music, demonstrating how different architectures adapt to the unique characteristics of these genres, including microtonal scales and intricate melodic phrasing.
\end{enumerate}
The remainder of this paper is structured as follows: Section~\ref{Adapters-Discussion} provides a discussion on the general theory regarding the trade-offs between music quality and efficiency in adapter-based fine-tuning, along with an exploration of various adapter architecture choices and placement strategies. Section~\ref{experimental_setup} details the experimental setup, including the training and test datasets, the text prompts used, and the evaluation metrics. Section~\ref{result} presents the results and offers an in-depth analysis based on both metrics and human evaluations. Finally, Section~\ref{conclusion} concludes the paper with a summary of our findings and suggestions for future research directions.

\section{Parameter-Efficient Fine-Tuning with Adapters}
\label{Adapters-Discussion}
In this section, we discuss the design considerations for integrating adapters into music generation models. While adapter-based Parameter-Efficient Fine-Tuning (PEFT) techniques offer significant computational benefits, they also introduce trade-offs that need to be balanced, primarily between quality and parameter efficiency. Specifically, the challenges in music generation, such as capturing long-range dependencies, multi-scale structures, and genre-specific features add complexity to this optimization.

\subsection{Model Selection}
We employ \textit{MusicGen} \cite{c:23} and \textit{Mustango} \cite{melechovsky-etal-2024-mustango} to explore cross-genre adaptation in music audio generation. \textit{MusicGen} is a Transformer-based autoregressive model that can generate music from text prompts and from partial melodies being prompted using EnCodec representations, excelling in style transfer and offering high controllability. In contrast, \textit{Mustango} extends pre-trained language models for text-to-music synthesis by leveraging a diffusion model, which conditions on text prompts, chord progressions, and beat information. This distinction allows us to analyze how different architectures adapt to underrepresented non-Western genres, providing valuable insights into adapter placement and training methodologies for distinct model types.

\subsection{Trade-offs between Efficiency and Quality}

Fine-tuning large-scale music generation models like \textit{MusicGen} and \textit{Mustango} traditionally requires updating billions of parameters, which is both computationally expensive and time-consuming. In contrast, adapter-based fine-tuning modifies only a small subset of model parameters, leaving the rest of the pre-trained weights frozen. This enables significant reductions in memory usage and computational cost, making fine-tuning feasible even in low-resource settings.

However, the main challenge with adapter-based PEFT lies in optimizing the trade-off between parameter efficiency and generation quality:

\begin{itemize} \item \textbf{Efficiency:} Adapters drastically reduce the number of parameters that need to be fine-tuned, allowing for faster training and lower computational requirements. In our experiments, we explore adapter sizes ranging from 2M to 70M parameters. By limiting the fine-tuning to a small portion of the model, ranging from 0.1\% to 5\% of the base model, adapters help make large-scale models more accessible for tasks requiring fewer computational resources. \item \textbf{Quality:} While adapters improve efficiency, they may limit the model’s ability to generate complex, high-quality music. Because adapters only adjust intermediate layers, they may not capture the full depth of long-range dependencies or intricate musical structures like multi-instrument compositions or evolving melodic lines. Full fine-tuning, which updates all parameters, allows for richer model expressiveness but comes at a higher computational cost. \end{itemize}

Thus, the goal is to find the optimal balance between parameter efficiency (minimizing the number of parameters updated) and quality (maintaining the model's ability to generate high-quality, complex music). The right configuration will depend on the task and available resources, and our experiments aim to explore this balance in detail.

\subsection{Adapter Architecture Design Choices}
Music generation poses unique challenges that influence how adapters should be designed and placed within the model. For Hindustani Classical \cite{jairazbhoy1971rāgs} and Maqam \cite{signell2008makam} music, adapter-based fine-tuning must address unique challenges such as complex melodic progressions, microtonal nuances, and long-form structures. Fully connected layer-based adapters offer efficiency but may struggle to capture the local temporal or hierarchical dependencies inherent in specific musical genres or traditions, making CNNs a better choice for these tasks. Convolution (CNN)-based adapters can model short-term dependencies effectively, capturing local patterns in music. However, they struggle to represent extended melodic developments, which require long-range dependencies that are vital in many musical traditions. In contrast, Transformer-based adapters are designed to capture long-term dependencies by leveraging self-attention mechanisms that allow them to model relationships across the entire sequence. This makes Transformers more effective than CNNs for capturing extended melodic phrasing and intricate, evolving musical structures, which are crucial in complex compositions. However, they are more computationally resource-intensive during training and require more data for training and adaptation. We experiment with three different adapter architectures, with Figure~\ref{fig:architecture} illustrating the key components of each adapter type.

\begin{itemize}
    \item \textbf{Fully connected layer-based:} Fully connected layers \cite{houlsby2019parameter} serve as efficient feature extractors, enabling compact representations of musical sequences. However, they are not as good as transformers and cnn architectures in capturing long-range dependencies and time sequence modelling which are crucial for raga and maqam progression. In our setup, this adapter compresses the input sequence into a low-dimensional bottleneck space using a down-projection layer, applies GELU \cite{hendrycks2016gaussian} activation, and restores the original sequence length through an up-projection layer. The choice of the bottleneck dimension plays a key role in determining the trainable parameter count, as a larger bottleneck would increase the number of parameters in both projection layers. A dropout layer prevents overfitting, and residual connections ensure the retention of key musical features. For \textit{MusicGen}, to maintain compatibility with the stereo (2-channel) format, the output is expanded accordingly. For \textit{Mustango}, this adapter type is not used, as it follows a UNet architecture with inputs having more than three dimensions, making only CNN and Transformer architectures suitable for use.
    \item \textbf{Convolution-Based (CNN):} Convolution neural network \cite{10.5555/303568.303704} adapters are effective in capturing local dependencies and fine-grained features which can include gamakas\footnote{Gamaka can be understood as embellishment done on a note or between two notes.}, meends\footnote{In Hindustani music, meend refers to a glide from one note to another.}, and murkis\footnote{Murki is a short taan or inverted mordent in Hindustani classical music.} key ornamentations in Hindustani Classical and Maqam music. In our setup for MusicGen, the adapter begins with a down-projection convolutional layer that reduces the input dimensionality while preserving sequence length. A deep residual bottleneck module follows, consisting of residual blocks \cite{yu2017dilated} that enable the model to process both short-term and long-term dependencies in the music signal. The bottleneck dimension directly influences the number of trainable parameters, particularly in convolutional layers and residual blocks. Additionally, a Squeeze-and-Excitation (SE) block is added in MusicGen which \cite{hu2018squeeze} applies channel-wise weight, allowing the adapter to dynamically focus on salient musical features. The up-projection layer restores the original sequence dimensions, ensuring minimal disruption to the pre-trained model. A dropout layer is incorporated to prevent overfitting. 
    \item \textbf{Transformer-Based:} Transformer \cite{NIPS2017_3f5ee243} adapters excel in modeling long-range dependencies, making them ideal for capturing the extended melodic line, rhythm patterns, and other long-range dependencies in Hindustani Classical music, as well as the maqam modulations in Arabic and Turkish traditions. In our setup, this adapter begins with a down-projection layer that compresses the input into a lower-dimensional bottleneck space, which significantly influences the overall parameter count in both self-attention and feed-forward layers. A multi-head self-attention module follows, allowing the model to capture global relationships across the sequence, crucial for preserving musical structure and coherence in long-form compositions \cite{NIPS2017_3f5ee243}. Layer normalization is applied before and after the attention mechanism to stabilize training. A feed-forward network (FFN) with GELU \cite{hendrycks2016gaussian} activation refines the learned representations, followed by an up-projection layer that restores the original sequence length. Residual connections and dropout regularization ensure robustness and prevent overfitting. This architecture is consistent for both \textit{MusicGen} and \textit{Mustango}, with the key difference being that \textit{Mustango} uses a 2D Transformer architecture to accommodate its unique input structure.
\end{itemize}
\subsection{Placement Strategies for Adapters}
In our placement experiments, shown in Figure~\ref{fig:architecture}, we tested inserting adapters into both the middle and late layers of \textit{MusicGen}. For \textit{Mustango}, we performed similar experiments on intermediate vs final layer placement and UNet block-wise addition of adapters. For each configuration, we listened to the generated audio and evaluated it using standard metrics. We trained these models for more than 10 epochs, listening to the quality of the generated audio to assess the output produced by the model.

\subsubsection{MusicGen}
For \textit{MusicGen}, we found that placing adapters in the middle layers often led to a complete breakdown in generation. The audio was not just of low quality; it was severely distorted, often consisting of loud beeping sounds or static, with no recognizable musical structure. This failure occurred even for prompts where the original base models could generate coherent, high-quality music. We hypothesize that such issues arise due to lack of data to finetune these layers in comparison to the data used for pre-training.

These failures were also reflected quantitatively: both \textit{Fréchet Audio Distance (FAD)} and \textit{Fréchet Distance (FD)} values were significantly higher compared to late-layer placements, confirming that the outputs diverged sharply from realistic music distributions. In contrast, when adapters were placed only in the final layers, generation quality improved noticeably. The models retained their internal musical structure and were able to add stylistic details like ornamentation or microtonal phrasing without losing coherence. Subjective listening evaluations also favored these late-layer configurations, describing the outputs as more fluid, stylistically accurate, and less artifact-prone.

\begin{figure*}[!t]
    \centering
\includegraphics[width=1.05\linewidth,height=20cm, keepaspectratio]{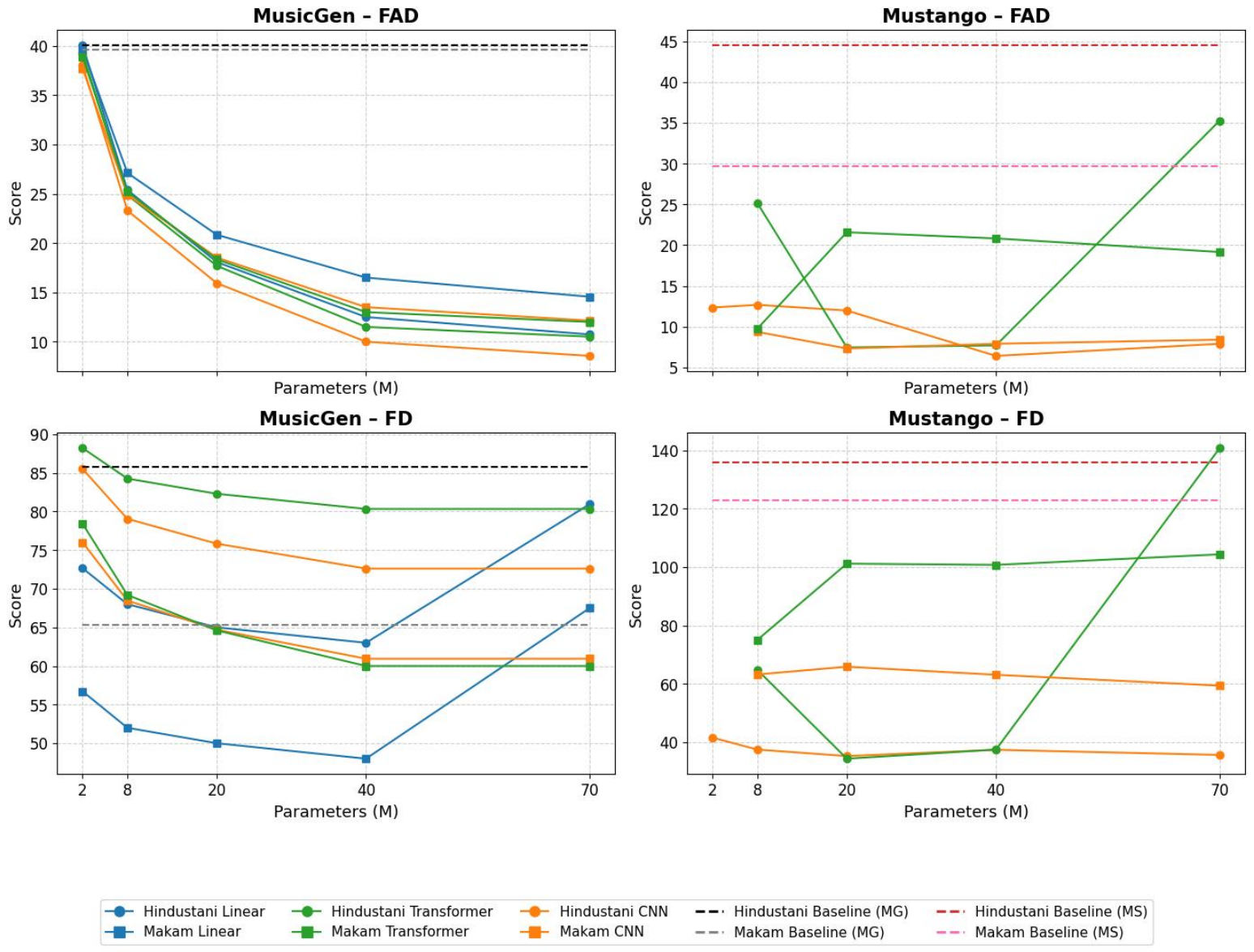}
    \caption{Comparing FAD and FD scores for \textit{MusicGen} \& \textit{Mustango} across three adapter architectures at varying parameter scales for Hindustani Classical and Turkish Makam music.}
    \label{fig:Adapter Scaling and Performance Trade-offs}
\end{figure*}

\subsubsection{Mustango}

For \textit{Mustango}, we incorporated adapters into the UNet architecture used for denoising, which consists of three main blocks: downsampling, mid-sampling, and upsampling. Each block contains multiple ResNet and Transformer layers. We experimented with various adapter placements after each transformer layer, after each ResNet layer, and after each complete block.

Our initial experiments showed that inserting adapters after individual Transformer layers resulted in outputs that lacked structure and often resembled random noise. Further analysis revealed that placing adapters after each complete block was far more effective. This configuration preserved the internal flow of information while allowing the model to adapt meaningfully to the target genre.

This pattern held consistently across both MusicGen and Mustango, and for both Hindustani Classical and Turkish Makam music. Inserting adapters into intermediate layers—such as within the encoder/decoder of MusicGen or within the internal layers of Mustango—led to degraded outputs. We hypothesize that these layers encode foundational musical concepts like timbre, harmonic progression, and rhythm. Modifying these activations disrupts the model's core understanding, causing it to forget or corrupt previously learned representations.

Finally, we observed that removing adapters from any block distorted learning, suggesting that adapter presence at the end of each block—downsampling, mid-sampling, and upsampling—is crucial for stability. These findings underscore the importance of adapter placement and highlight the value of late-layer interventions in preserving structure while enabling stylistic adaptation.

%\begin{table}[!t]
%  \centering
%  \begin{tabular}{|l|c|c|c|c|c|}
%    \hline
%    Model & 2M & 8M & 20M & 40M & 70M \\
%    \hline
%    MG-P & 10 & 12 & 19 & 27 & 38 \\
%    MG-C & 7 & 9 & 12 & 20 & 26 \\
%    MG-T & 8 & 10 & 15 & 22 & 30 \\
%    MS-P & X & Y & Z & A & B \\
%    MS-C & 12 & Y & Z & A & B \\
%    MS-T & X & Y & Z & A & B \\
%    \hline
%  \end{tabular}
%  \caption{GPU hours used across six adapter models at different parameter configurations. Values are rounded up to the nearest whole number.}
%  \label{tab:ad_placement}
%\end{table}

\section{Experimental Setup}
\label{experimental_setup}
For our adapter experiments, we chose two distinct non-Western musical genres, Hindustani Classical~\cite{jairazbhoy1971rāgs} and Turkish Makam~\cite{signell2008makam}, both of which are significantly underrepresented in music generation research and datasets. Additionally, we utilized two open-source models: \textit{MusicGen}~\cite{c:23} and \textit{Mustango}~\cite{melechovsky-etal-2024-mustango}. Our study begins with an overview of dataset creation, followed by model selection, adapter architectures, and, finally, the training process and evaluation metrics.
\subsection{Dataset}

\subsubsection{Dataset Selection}
Our research required a broad collection of non-Western music accompanied by detailed metadata, leading us to select the Dunya corpus \cite{porter-2013}, a key resource within the CompMusic initiative \cite{serra2014creating}. This dataset encompasses more than 1,300 hours of recordings across various non-Western traditions, including Carnatic, Hindustani, Turkish Makam, Beijing Opera, and Arab Andalusian music. We concentrated on Hindustani Classical and Turkish Makam due to their intricate and similar melodic and rhythmic frameworks, which significantly differ from Western musical structures. For Hindustani Classical, we cumulated 329.16 hours of labeled audio. Similarly, for Turkish Makam, we retrieved metadata and sample recordings via the Dunya dataset API, amassing 269.71 hours of content, leading to a total of approximately 600 hours of total data.

To maintain uniformity and optimize computational performance, we processed the dataset by shortening longer recordings into 30-second segments while preserving all metadata. These metadata elements, rich in genre-specific characteristics, were embedded within prompt templates for model training. Furthermore, we adjusted the audio sampling rate to align with model specifications: 32 kHz for \textit{MusicGen} and 16 kHz for \textit{Mustango}. The dataset was divided into training (80\%) and testing (20\%) subsets, ensuring that audio clips in the test set originated from different songs than those in the training set to prevent distributional overlap. After preprocessing, we retained 246.87 hours of Hindustani Classical and 202.28 hours of Turkish Makam music, with 208.58 hours and 157.01 hours, respectively, allocated for training.

\subsubsection{Prompt Formation}
The metadata from the dataset provides genre-specific information for each audio clip, including three key details critical to our study: melodic line, rhythmic pattern, and instrumentation. For the
melodic line, we extracted the raga (a melodic framework in Hindustani Classical music) and Makam (a system of melodic modes in Turkish music). For rhythmic patterns, we identified taal (rhythm structure) in Indian music and usul (a sequence of rhythmic strokes) in Turkish music. Additionally, we extracted the metadata for the instruments (including voice) played in each audio sample. 

The Hindustani Classical dataset includes 21 different instrument types, such as the Pakhavaj, Zither, Sarangi, Ghatam, Harmonium, and Santoor, along with vocals. It spans 200 ragas and 26 distinct taals. The Turkish Makam dataset features 42 makam-specific instruments, such as Oud, Tanbur, Ney, Davul, Clarinet, Kös, Kudüm, Yaylı Tanbur, Tef, Kanun, Zurna, Bendir, Darbuka, Classical Kemençe, Rebab, Çevgen, and vocals. It encompasses 100 different makams and 62 distinct usuls.

\begin{figure*}[!t]
    \centering
    \includegraphics[height=0.22\textwidth]
{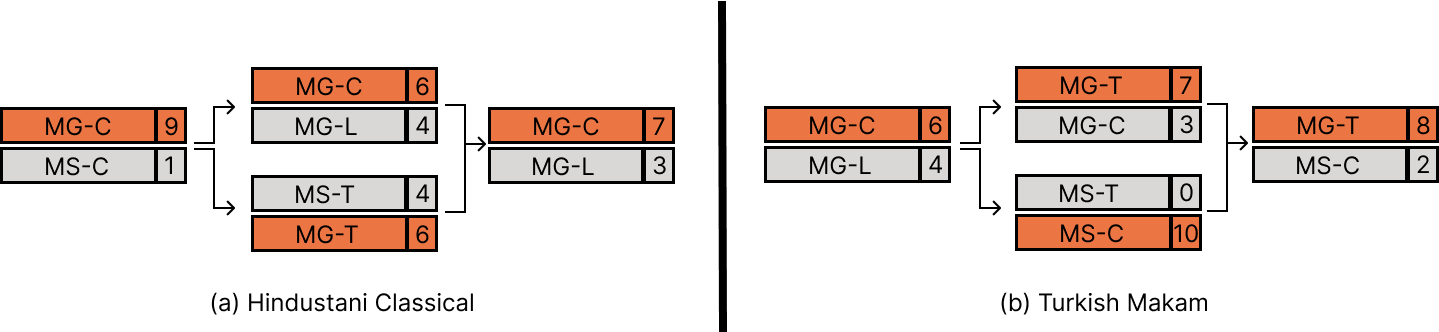}
    \caption{Human evaluation of subjective quality and aesthetics for (a) Hindustani Classical and (b) Turkish Makam music.}
    \label{fig:combined_subjective}
\end{figure*}

For training text-to-music models, we have to generate text prompts for each equivalent audio sample. To make sure each prompt is descriptive, we utilise the above metadata combinations for each audio sample and fit them into five pre-defined prompt templates. Each prompt is a paraphrased version of the same base, with blanks for melodic line, instruments, and rhythmic pattern. This augmentation promotes diversity and helps prevent overfitting. Unlike the pretraining prompts used by MusicGen and Mustango(shown in Table \ref{tab:prompt_compact})—which are largely Western-centric and lack fine-grained metadata—our prompts explicitly specify genre, melodic structure, instrumentation, and rhythm tailored to Hindustani Classical and Turkish Makam.

\begin{table}[h!]
\centering
\begin{tabular}{|l|p{4.6cm}|}
\hline
\textbf{Setting} & \textbf{Example Prompt} \\ \hline
\textbf{Pretraining-MusicGen} & “80s pop track with bassy drums and synth” \\ \hline
\textbf{Pretraining-Mustango} & “Western instrumental with electric guitar, rim shots, and E major key” \\ \hline
\textbf{Ours} & “Hindustani Classical with Tanpura, Tabla, Voice, Harmonium in Bhairavi raga in Teental.” \\ \hline
\end{tabular}
\caption{Prompt styles across MusicGen, Mustango, and our approach.}
\label{tab:prompt_compact}
\end{table}

\subsection{Training Setup}
For fine-tuning \textit{MusicGen}, we utilized two RTX A6000 GPUs, while for \textit{Mustango}, we employed a single RTX A6000 GPU each with a capacity of 48 GB. We provide detailed information on training time, inference, and GPU memory usage in the subsequent sections. In the case of \textit{MusicGen}, adapter blocks were fine-tuned using the AdamW~\cite{Loshchilov2017DecoupledWD} optimizer with a learning rate of 5e-5 and a weight decay of 0.05, leveraging an MSE-based reconstruction loss. Similarly, for \textit{Mustango}, the adapter block was fine-tuned using AdamW with a learning rate of 4.5e-5 and a weight decay of 0.0001, also employing MSE-based reconstruction loss. While \textit{Mustango} reaches optimal performance in 10-15 epochs, \textit{MusicGen} takes 18-25 epochs to train, with the optimal number of epochs determined using early-stopping \cite{prechelt2002early} on the validation set. The training dataset was further partitioned into 90\% for training and 10\% for validation to ensure a balanced evaluation during fine-tuning. 

\subsection{Evaluation Metrics}
\subsubsection{Objective Evaluation}
We extract 400 audio clips from the test set to construct our evaluation prompt corpus. To quantify the similarity between the generated music and the reference test corpus, we calculate Fréchet Audio Distance (FAD) \cite{Kilgour2019FrchetAD}, Fréchet Distance (FD). The implementation of these metrics is carried out using the AudioLDM~\cite{liu2023audioldm} framework. To compute the distributions required for FAD, FD, we employ PANN-CNN14~\cite{Kong_Cao_Iqbal_Wang_Wang_Plumbley_2020} as the feature extraction backbone for processing each audio sample. 

\subsubsection{Subjective Evaluation}
For subjective human evaluation, we used a similar framework  as described in ~\citet{mehta2025music} by implementing an arena-style setup and an elimination-based strategy to compare the generative performance of our models subjectively. In this approach, for each of the 10 distinct prompts, we generated audio snippets from two models at a time and compared them on the basis of clarity, auditory pleasantness, and freedom from unwanted artifacts. The model that wins on a majority of these prompt-wise comparisons advances to the next round, and this process is repeated until a single model remains. We compare five models (three from \textit{MusicGen} and two from \textit{Mustango}) for both Hindustani and Makam, resulting in 40 total matchups per genre. Ultimately, the best-performing model is identified as the winner for that genre. For subjective evaluations, the annotators included two people--one music expert and another avid listener who annotated both genres. Due to resource constraints, we limited the number of annotators. Given that our evaluation focused on overall quality rather than a detailed analysis of specific components like melodic line, rhythm, and timbre, we found that two annotators were sufficient.

\section{Results and Discussion}
\label{result}

We evaluated three adapter architectures at multiple parameter scales (ranging from about 2M up to 70M) in two distinct music genres: Hindustani Classical and Turkish Makam. Specifically, for \textit{MusicGen}, we tested \textit{MusicGen}-Linear (MG-L), \textit{MusicGen}-CNN (MG-C), and \textit{MusicGen}-Transformer (MG-T), while for \textit{Mustango}, we examined \textit{Mustango}-Convolutions (MS-C), and \textit{Mustango}-Transformer (MS-T). A linear layer-based adapter can be used for \textit{MusicGen} but not for \textit{Mustango} due to fundamental differences in how these models process data. \textit{MusicGen} utilizes discrete token representations from EnCodec within a Transformer-based architecture, allowing linear layer adapters to efficiently modify token embeddings without disrupting structure. In contrast, \textit{Mustango} is a diffusion-based model that operates on continuous 3D latent representations, where all transformations occur channel-wise rather than at the token level. Since MLPs require flattened 2D inputs, they cannot properly process \textit{Mustango’s} structured latent space.

\subsection{Objective Evaluations}
\label{sec:objective}
Figure~\ref{fig:Adapter Scaling and Performance Trade-offs} (left plots) shows results for \textit{MusicGen} for both Hindustani Classical and Turkish Makam music. Across all configurations, we find that the 40M parameter scale offers the best FAD and FD scores ensuring optimal trade-off between music generation quality and the adapter size for the given amount of data. At this scale, FAD and FD scores stabilize at low values across architectures and genres, while training time remains significantly lower than that of the larger 70M models. For example, Hindustani\_MG-C and Makam\_MG-T at 40M both achieve excellent FAD scores (10.0 and 13.0, respectively), indicating good audio quality, without incurring the diminishing returns observed beyond this scale. This sweet spot can be attributed to the nature of adapter-based PEFT; small adapters (e.g., 2M or 8M) lack sufficient capacity to capture complex musical dependencies, conversely, very large adapters (e.g., 70M) not only increase memory and compute requirements but also risk overfitting or disrupting the model's learned representations particularly noticeable in MG-L configurations, where FD scores for both genres degrade sharply at 70M. Among architectures, convolution-based adapters (MG-C) performs best for Hindustani classical, while transformer-based adapters (MG-T) perform best for Turkish Makam.

Figure~\ref{fig:Adapter Scaling and Performance Trade-offs} (right plots) shows \textit{Mustango} results for both Hindustani Classical and Turkish Makam music. Unlike \textit{MusicGen}, whose best trade-off emerged at the 40M scale \textit{Mustango} demonstrates its most stable performance in the 20M--40M parameter range. At smaller adapter sizes (2M or 8M), model capacity is insufficient for capturing complex musical structures, reflected in higher FAD scores (e.g., \texttt{Hindustani\_MS-C} at 2M hovers near 12). In contrast, the largest 70M adapters yield inconsistent improvements and can even degrade fidelity metrics (e.g., FD for \texttt{Makam\_MS-T} climbs above 100 at 70M), suggesting a risk of overfitting or destabilizing previously learned representations. In particular, 40M adapters typically achieve low FAD values, especially for Hindustani CNN (\texttt{MS-C}) at 6.4 and Makam CNN (\texttt{MS-C}) at 8.39.  Among architectures for \textit{Mustango}, convolution-based adapters (MS-C) performs best for both Hindustani classical and Turkish Makam with transformer based adapter (MS-T) matching the performance for Hindustani Classical but not for Turkish Makam.

Both \textit{MusicGen} and \textit{Mustango} share the common trend that \emph{mid-range adapter sizes} are optimal for the trade-offs of too-little capacity (leading to higher FAD/FD scores) and too-large capacity (increased risk of overfitting and larger sizes leading to heavier compute demands). In \textit{MusicGen}, the optimal point holds firmly at 40M; in \textit{Mustango}, peak performance is often achieved slightly earlier, around 20M parameters, though 40M remains competitive. In both models, architecture preferences are genre-dependent, with CNN adapters often favored for Hindustani classical and Transformer adapters for Turkish Makam.

\subsection{Subjective Evaluations}
\label{sec:subjective}
To complement our objective metrics, we conducted a comprehensive subjective evaluation of musical quality, focusing exclusively on the 40M parameter-scale adapters identified as optimal in Section~\ref{sec:objective}. Figure~\ref{fig:combined_subjective}(a) presents the results for Hindustani Classical music, where the convolution-based \textit{MusicGen} adapter (MG-C) emerged as the most preferred configuration, winning the majority of arena-style matchups. Annotators highlighted \textbf{MG-C’s notable clarity and coherence, especially when compared to the transformer-based counterpart MG-T}, which was criticized for its redundant musical phrases and reduced creative variation, ultimately leading to poorer aesthetic appeal in this genre.

The Mustango-CNN (MS-C) model also performed strongly, earning praise for its rich instrumental textures, faithful rendering of vocal ornamentations, and accurate adherence to prompt-specific details. However, its performance was hindered by a lack of structural clarity and audio instability—with frequent note misalignments and pitch inaccuracies, resulting in perceptibly lower audio quality despite its otherwise expressive output. We hypothesize that since \textit{Mustango} is pre-trained on a huge corpus of data, which has information about chord progressions and rhythm patterns used for conditioning the model. \textbf{When we adapt \textit{Mustango} to a new genre with only the text input without any chord or beat conditioning, it fails to properly align and stabilise the melody and rhythm.}

For Turkish Makam (Figure~\ref{fig:combined_subjective}(b)), the MusicGen-Transformer (MG-T) adapter was most favored. Annotators commended its ability to preserve long-form structure and navigate complex modal transitions, lending a natural and flowing character to the music. While MS-C remained a close contender in this genre as well, its output was sometimes perceived as less expressive or slightly repetitive over extended durations which explains the fact that it has higher FAD \textit{MusicGen} models.

Across both genres, linear adapters (MG-L) consistently underperformed in subjective evaluations, aligning with objective metrics where they recorded higher FAD and FD scores, especially at larger parameter scales. \textbf{A key insight from our analysis is the contrast between objective and subjective performance, particularly in the case of Hindustani Classical music. Here, \textit{Mustango} models achieved lower FAD scores, indicating stronger adherence to prompt which implies that it is closer to the ground truth distribution in terms of entropy.} However, this did not translate to better human evaluations—annotators found that despite its creative variety, \textit{Mustango} often lacked note alignment, structural stability, and audio clarity, which ultimately impacted the perceived quality of the generated music. In case of diversity, we observed that the MusicGen model generations were highly repetitive in its generation and had very narrow set of instruments, rhythm types and melodies in its generations leading to our understanding that Mustango could generate music using a broader set of musical attributes including instruments, rhythms and melodies which led to a lower FAD score since it is closer to the ground truth distribution. In contrast, MusicGen-CNN (MG-C), though slightly less diverse, was rated higher for coherence, clarity, and musicality explaining the difference in objective and subjective evaluations.

\subsection{Computational Efficiency}
The computational efficiency of music generation models varies significantly depending on the underlying architecture and overall parameter count. In \textit{MusicGen}, the authors note that larger base models (e.g., 300M, 1.5B, and 3.3B parameters) deliver improved performance but incur correspondingly higher training and inference times. However, exact training durations for these large architectures were not reported. By contrast, \textit{Mustango} was trained on 4 Nvidia Tesla V100s and 8 Quadro RTX 8000s, taking approximately 5-10 days with an effective batch size of 32.

\begin{figure}[!t]
    \centering
    \includegraphics[width=0.48\textwidth]{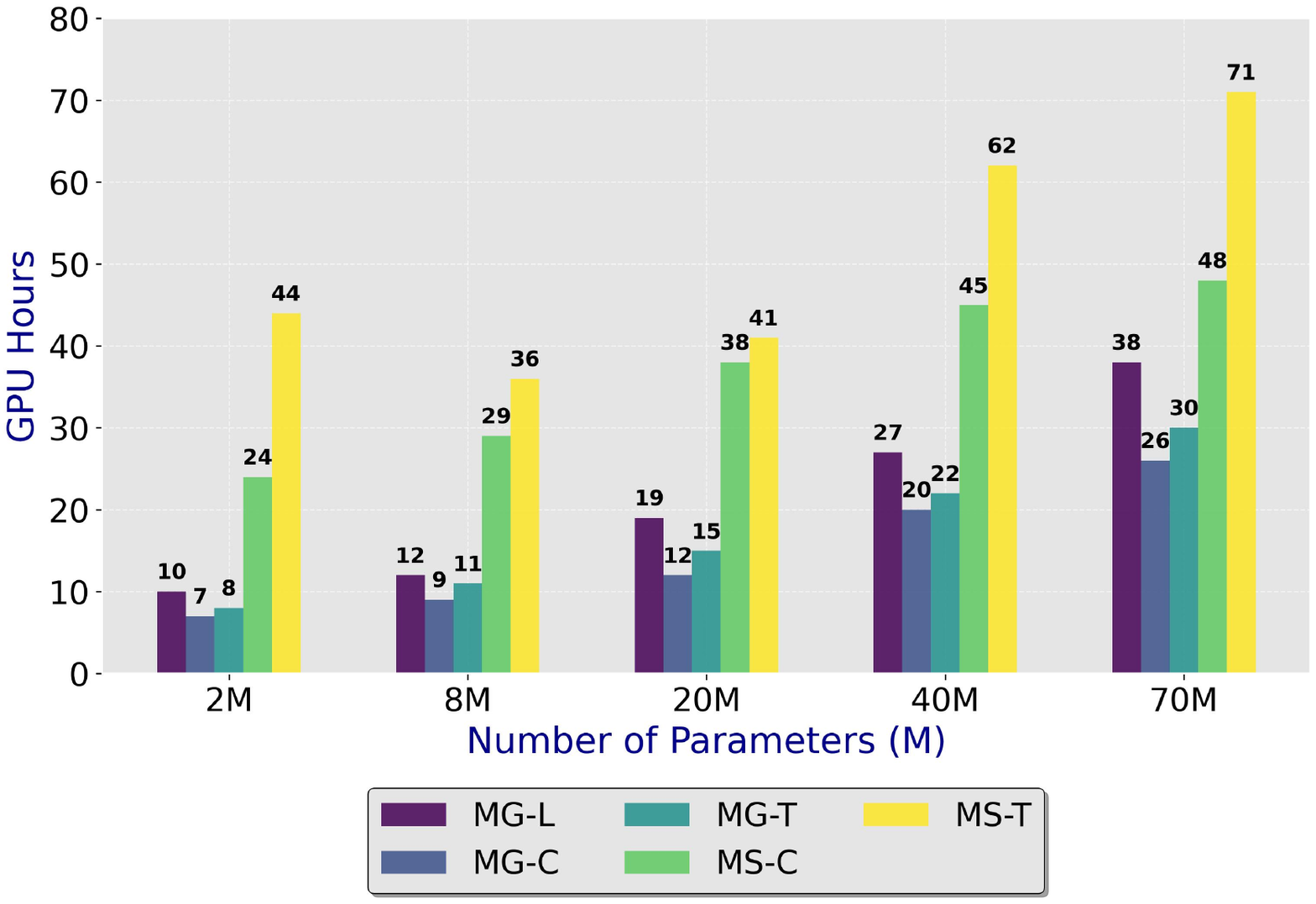}
    \caption{GPU hours used across five adapter models at different parameter configurations. Values are rounded up to the nearest whole number.}
    \label{fig:gpu_hours}
\end{figure}

Our study focuses on adapter-based parameter-efficient fine-tuning, whose computational demands are comparatively much lower. As shown in Figure~\ref{fig:gpu_hours}, even for our 70M-parameter adapters \textit{MusicGen} (denoted \texttt{MG-} in the figure) require at most 38 hours for \texttt{MG-L}, 26 hours for \texttt{MG-C}, and 30 hours for \texttt{MG-T}, whereas \textit{Mustango} adapters (denoted \texttt{MS-} in the figure) require between 48 and 71 hours for \texttt{MG-C} and \texttt{MG-T} respectively under smaller batch sizes of 4. These results demonstrate that parameter-efficient approaches can sharply reduce training time while maintaining strong generative performance.

Focusing on our best-performing 40M-scale models (as identified in Sections~\ref{sec:objective} and~\ref{sec:subjective}), we observe that:
\begin{itemize}
    \item \textbf{\textit{MusicGen}} shows excellent efficiency in its top-performing adapters: for Hindustani Classical, the 40M Convolution (\texttt{MG-C}) configuration takes only 20 hours of GPU time, while for Turkish Makam, the 40M Transformer (\texttt{MG-T}) configuration takes 22 hours. For MusicGen configurations, at inference time, generating each sample requires approximately 3 seconds with 40GB of GPU memory.
    \item \textbf{\textit{Mustango}}, which benefits most from a Convolution-based adapter for both Hindustani and Makam at the 40M scale (\texttt{MS-C}), requires about 45 hours of GPU time. Although longer than its \textit{MusicGen} counterpart, it is still substantially more efficient than training a large model from scratch. At inference time, Mustango requires 100 seconds (depending on the number of denoising steps which in our case was 200) for generating 10 second audios with a batch size of 4 and GPU memory of 32GB on a single GPU.
\end{itemize}

Overall, these results confirm that tailoring adapter size and architecture can achieve a favorable trade-off between computation and generation quality. Even though \textit{Mustango} adapters require more hours in part due to the diffusion-based architecture, the overall time remains considerably lower than what would be necessary for end-to-end fine-tuning of large-scale music generation frameworks.

% add the compute power required

\section{Conclusion and Future Work}
\label{conclusion}
In this study, we systematically explored the impact of different adapter configurations—linear, convolution-based, and transformer-based—on parameter-efficient fine-tuning (PEFT) for music generation models. Using controlled experiments on \textit{MusicGen} and \textit{Mustango}, we evaluated their performance across two culturally rich genres: Hindustani Classical and Turkish Makam.

Overall, \textit{MusicGen} produced higher quality audio while taking lesser training time, with transformer and CNN adapters showing complementary strengths across genres—the CNN adapter excelled in Hindustani Classical, while the transformer adapter was more effective in Turkish Makam. For \textit{Mustango}, the CNN-based adapter matched or outperformed the transformer variant in both objective and subjective evaluations.

Qualitative analysis further revealed that \textit{Mustango} outputs exhibited greater diversity and better adherence to prompts in Hindustani Classical, while \textit{MusicGen} outputs were more homogeneous but rated higher in quality due to their clarity and coherence. In the case of Turkish Makam, \textit{Mustango} showed high FAD scores, reflecting poor alignment with reference distributions, and subjective feedback also pointed to redundant and less expressive outputs.

Our results indicate that the 40M parameter scale is well-suited for the dataset sizes considered, though optimal adapter configurations may vary with larger or more complex datasets. While our study offers key insights into adapter-based PEFT for music generation, several promising directions remain:

\begin{itemize}
    \item \textbf{Hybrid PEFT Methods:} Combining adapters with techniques like LoRA \cite{hu2022lora} or prefix tuning \cite{li2021prefix} may enhance efficiency and adaptability.
    
    \item \textbf{Cultural Extension:} Applying our approach to traditions like Carnatic \cite{vijayakrishnan2007grammar}, Persian Dastgah \cite{nettl2001music}, or Gamelan \cite{becker1993gamelan} can test generalizability across diverse genres.
    
    \item \textbf{Cross-Architecture Transfer:} Exploring whether adapters trained on transformer-based models (e.g., \textit{MusicGen}) can be transferred to diffusion-based ones (e.g., \textit{Mustango}) \cite{yang2024cross}.
    
    \item \textbf{Scaling Trade-offs:} Inspired by Chinchilla \cite{hoffmann2022chinchilla} and scaling laws \cite{kaplan2020scaling}, future work can investigate the interplay between model size, data volume, and genre diversity.
\end{itemize}

By addressing these directions, future work can further refine adapter-based fine-tuning, making music AI more accessible, efficient, and expressive, while expanding its applicability to a broader range of musical styles and generative tasks.
% For BibTeX users:
\bibliographystyle{ACM-Reference-Format}
\balance
\bibliography{main}

% For non BibTeX users:
%\begin{thebibliography}{citations}
% \bibitem{Author:17}
% E.~Author and B.~Authour, ``The title of the conference paper,'' in {\em Proc.
% of the Int. Society for Music Information Retrieval Conf.}, (Suzhou, China),
% pp.~111--117, 2017.
%
% \bibitem{Someone:10}
% A.~Someone, B.~Someone, and C.~Someone, ``The title of the journal paper,''
%  {\em Journal of New Music Research}, vol.~A, pp.~111--222, September 2010.
%
% \bibitem{Person:20}
% O.~Person, {\em Title of the Book}.
% \newblock Montr\'{e}al, Canada: McGill-Queen's University Press, 2021.
%
% \bibitem{Person:09}
% F.~Person and S.~Person, ``Title of a chapter this book,'' in {\em A Book
% Containing Delightful Chapters} (A.~G. Editor, ed.), pp.~58--102, Tokyo,
% Japan: The Publisher, 2009.
%
%\end{thebibliography}

\end{document}